\documentclass[prd,superscriptaddress,twocolumn,nofootinbib,showpacs,preprintnumbers,floatfix]{revtex4}

\usepackage{mathrsfs}
\usepackage{amsfonts,amssymb,amsmath}
\usepackage{graphicx,epsfig}
\usepackage{multirow}

\begin{document}

\title{Supersymmetric Dark Matter at XENON100 and the LHC:
\\No-Scale $\cal{F}$-$SU(5)$ Stringy Correlations}

\author{Tianjun Li}

\affiliation{State Key Laboratory of Theoretical Physics, Institute of Theoretical Physics,
Chinese Academy of Sciences, Beijing 100190, P. R. China }

\affiliation{George P. and Cynthia W. Mitchell Institute for Fundamental Physics and Astronomy,
Texas A$\&$M University, College Station, TX 77843, USA }

\author{James A. Maxin}

\affiliation{George P. and Cynthia W. Mitchell Institute for Fundamental Physics and Astronomy,
Texas A$\&$M University, College Station, TX 77843, USA }

\author{Dimitri V. Nanopoulos}

\thanks{\textit{\textbf{\normalsize Contribution to the Proceedings of UCLA Dark Matter 2012, Marina del Rey, CA, 22-24 February 2012, based on a talk given by Dimitri V. Nanopoulos.}}}

\affiliation{George P. and Cynthia W. Mitchell Institute for Fundamental Physics and Astronomy,
Texas A$\&$M University, College Station, TX 77843, USA }

\affiliation{Astroparticle Physics Group, Houston Advanced Research Center (HARC),
Mitchell Campus, Woodlands, TX 77381, USA}

\affiliation{Academy of Athens, Division of Natural Sciences,
28 Panepistimiou Avenue, Athens 10679, Greece }

\author{Joel W. Walker}

\affiliation{Department of Physics, Sam Houston State University,
Huntsville, TX 77341, USA }


\begin{abstract}

We complete an investigation of the observable signatures of No-Scale flipped $SU(5)\times U(1)_X$ grand unified theory
with TeV-scale vector-like particles (No-Scale $\cal{F}$-$SU(5)$) at the LHC and dark matter direct detection experiments.  We feature a dark matter candidate which is over 99\% bino due to a comparatively large Higgs bilinear mass $\mu$ term around the electroweak scale, and hence
automatically satisfy the present constraints from the XENON100 and CDMS/EDELWEISS experiments.
We do however expect that the continued XENON100 run and extension to 1-ton may begin to probe our model.
Similarly, our model is also currently under probe by the LHC through a search for events with ultra-high multiplicity hadronic jets,
which are a characteristic feature of the distinctive No-Scale $\cal{F}$-$SU(5)$ mass hierarchy.

\end{abstract}


\preprint{ACT-12-12, MIFPA-12-32}

\maketitle


The search for supersymmetry (SUSY) and dark matter at the Large Hadron Collider (LHC) has been progressing since March 2010, steadily accumulating data from ${\sqrt s}=7-8$ TeV proton-proton collisions by the CMS and ATLAS Experiments, and has reached an integrated luminosity of $15~{\rm fb}^{-1}$ to date, with possibly over $25~{\rm fb}^{-1}$ anticipated by the end of 2012. Nonetheless, early results have produced no definitive signal of supersymmetry or dark matter, drastically constraining the experimentally viable parameter space of the CMSSM and mSUGRA, in addition to the entire landscape of supersymmetric models. The lack of convincing evidence of supersymmetry thus far has increased the constraints on the viable CMSSM and mSUGRA model space, posing the question whether there exist SUSY and/or superstring post-Standard Model extensions that can elude the currently enforced LHC constraints, though surviving within the 2012 reach of the LHC.

The search for dark matter (DM) is being led on the direct detection front by the XENON100 collaboration~\cite{Aprile:2011hi}, 
whose expansion in the last year to a fiducial detector mass of 62 kg of ultra-pure
liquid xenon has promptly captured a near ten-fold improvement over 
the CDMS and EDELWEISS experiments~\cite{CDMS:2011gh} in the upper bound on the spin-independent
cross section for scattering WIMPs against nucleons.  This limit
likewise begins to cut incisively against the favored regions of the CMSSM.

The exploration of the existence of dark matter persists not only between parallel experimental search strategies, but also between
alternative theoretical proposals.  We have studied in comprehensive detail a promising model by the name of No-Scale 
$\cal{F}$-$SU(5)$~\cite{Li:2010ws, Li:2010mi,Li:2010uu,Li:2011dw, Li:2011hr, Maxin:2011hy, Li:2011xu,Li:2011gh,Li:2011rp,Li:2011fu,Li:2011xg,Li:2011ex,Li:2011av,Li:2011ab,Li:2012hm,Li:2012tr,Li:2012ix,Li:2012yd,Li:2012qv,Li:2012jf}, which 
is constructed from the merger of the ${\cal F}$-lipped $SU(5)$ Grand Unified Theory
(GUT)~\cite{Barr:1981qv,Derendinger:1983aj,Antoniadis:1987dx},
two pairs of hypothetical TeV scale vector-like supersymmetric multiplets with origins in
${\cal F}$-theory~\cite{Jiang:2006hf,Jiang:2009zza,Jiang:2009za,Li:2010dp,Li:2010rz},
and the dynamically established boundary conditions of No-Scale
Supergravity~\cite{Cremmer:1983bf,Ellis:1983sf, Ellis:1983ei, Ellis:1984bm, Lahanas:1986uc}.
The experimentally viable parameter space of No-Scale $\cal{F}$-$SU(5)$ has been comprehensively mapped~\cite{Li:2011xu}, which satisfies the
``bare minimal'' phenomenological constraints, possesses the correct cold DM (CDM) relic density, and is consistent
with a dynamic determination by the secondary minimization of the Higgs potential via the ``Super No-Scale''
mechanism~\cite{Li:2010uu,Li:2011dw,Li:2011xu,Li:2011ex}.

The No-Scale $\cal{F}$-$SU(5)$ construction inherits all of the most beneficial phenomenology~\cite{Nanopoulos:2002qk}
of flipped $SU(5)$~\cite{Barr:1981qv,Derendinger:1983aj,Antoniadis:1987dx}, as well as all of the valuable theoretical motivation of No-Scale
Supergravity~\cite{Cremmer:1983bf,Ellis:1983sf, Ellis:1983ei, Ellis:1984bm, Lahanas:1986uc},
including a deep connection to the string theory infrared limit
(via compactification of the weakly coupled heterotic theory~\cite{Witten:1985xb} or 
M-theory on $S^1/Z_2$ at the leading order~\cite{Li:1997sk}),
and a mechanism for SUSY breaking which preserves a vanishing cosmological constant at the tree level
(facilitating the observed longevity and cosmological flatness of our Universe~\cite{Cremmer:1983bf}).

Mass degenerate superpartners for the known SM fields have not been observed, therefore SUSY must itself be broken near the TeV scale.
In mSUGRA, this begins in a hidden sector, and the secondary propagation by gravitational interactions into the observable sector is parameterized by universal SUSY-breaking ``soft terms'' which include the gaugino mass $M_{1/2}$, scalar mass
$M_0$ and the trilinear coupling $A$.  The ratio of the low energy Higgs vacuum expectation values (VEVs) tan$\beta$, and the sign of
the SUSY-preserving Higgs bilinear mass term $\mu$ remain undetermined, while the magnitude of the $\mu$ term and its bilinear soft term $B_{\mu}$
are determined by the $Z$-boson mass $M_Z$ and tan$\beta$ after electroweak symmetry breaking (EWSB).  In the simplest No-Scale scenario,
$M_0$=A=$B_{\mu}$=0 at the unification boundary, while the entire set of low energy SUSY breaking soft-terms evolve down 
from a single non-zero parameter $M_{1/2}$. As a result, the particle spectrum is proportional to $M_{1/2}$ at leading order,
rendering the bulk ``internal'' physical properties invariant under an overall rescaling.

The (formerly negative) one-loop $\beta$-function
coefficient of the strong coupling $\alpha_3$ becomes precisely zero, flattening the RGE running, and generating a wide
gap between the large $\alpha_{32} \simeq \alpha_3(M_{\rm Z}) \simeq 0.11$ and the much smaller $\alpha_{\rm X}$ at the scale $M_{32}$ of the intermediate flipped $SU(5)$ unification of the $SU(3)_C \times SU(2)_{\rm L}$ subgroup.  This facilitates a very significant secondary running phase
up to the final $SU(5) \times U(1)_{\rm X}$ unification scale~\cite{Li:2010dp}, which may be elevated by 2-3 orders of magnitude
into adjacency with the Planck mass, where the $B_\mu = 0$ boundary condition fits well~\cite{Ellis:2001kg,Ellis:2010jb,Li:2010ws}. We denote this final $SU(5) \times U(1)_{\rm X}$ unification scale as $M_{\cal F}$, where for the experimentally viable parameter space, $M_{\cal F}$ transpires at about 4-6$\times 10^{17}$ GeV, right near the string scale of $\sim$5$\times 10^{17}$ GeV, providing a very natural solution to the ``little hierarchy'' problem.

The modifications to the $\beta$-function coefficients from introduction of the vector-like multiplets have a
comparable effect on the RGEs of the gauginos.  Specifically, the color-charged gaugino mass $M_{\rm 3}$ likewise evolves down from the
high energy boundary flat, obeying the relation $M_3/M_{1/2} \simeq \alpha_3(M_{\rm Z})/\alpha_3(M_{32}) \simeq \mathcal{O}\,(1)$,
which precipitates a conspicuously light gluino mass assignment.
The $SU(2)_{\rm L}$ and hypercharge $U(1)_{\rm Y}$ associated gaugino masses are by
contrast driven downward from the $M_{1/2}$ boundary value by roughly the ratio of their corresponding gauge couplings
$(\alpha_2,\alpha_{\rm Y})$ to the strong coupling $\alpha_{\rm s}$.
The large mass splitting expected from the heaviness of the top quark via its strong coupling to the Higgs (which is also key to generating an appreciable radiative Higgs mass shift $\Delta~m_h^2$~\cite{Okada:1990vk,Okada:1990gg,Haber:1990aw,Ellis:1990nz,Ellis:1991zd}) is responsible for a rather light stop squark $\widetilde{t}_1$. The distinctively predictive $m_{\tilde{t}_1} < m_{\tilde{g}} < m_{\tilde{q}}$ mass hierarchy of a light stop and gluino, both much lighter than all other squarks, is stable across the full No-Scale $\cal{F}$-$SU(5)$ model space, but is
not precisely replicated in any phenomenologically favored CMSSM constructions of which we are aware.

The spectrum associated with this mass hierarchy generates a unique event topology starting from the pair production of heavy squarks
$\widetilde{q} \widetilde{\overline{q}}$, except for the light stop, in the initial hard scattering process,
with each squark likely to yield a quark-gluino pair $\widetilde{q} \rightarrow q \widetilde{g}$.  Each gluino may be expected
to produce events with a high multiplicity of virtual stops or tops, via the (possibly off-shell) $\widetilde{g} \rightarrow \widetilde{t}_1 \overline{t}$ or $\widetilde{g} \rightarrow \widetilde{\overline{t}}_1 t$ transitions, which in turn may terminate into hard scattering products such as $\rightarrow W^{+}W^{-} b \overline{b} \widetilde{\chi}_1^{0}$ and $W^{-} b \overline{b} \tau^{+} \nu_{\tau} \widetilde{\chi}_1^{0}$, where the $W$ bosons will produce mostly hadronic jets and some leptons. The model described may then consistently produce a net product of eight or more jets emergent from a single squark pair production event, passing through a single intermediate gluino pair, resulting after fragmentation in an impressive signal of ultra-high multiplicity final state jet events.

The entirety of the viable $\cal{F}$-$SU(5)$ parameter space naturally features a dominantly bino LSP, at a purity greater than 99\%, as is exceedingly suitable for direct detection.  There exists no direct bino to wino mass mixing term. This distinctive and desirable model characteristic is guaranteed by the relative heaviness of the Higgs bilinear mass $\mu$, which in the present construction generically traces the the universal gaugino mass $M_{1/2}$ at the boundary scale $M_{\cal F}$, and subsequently transmutes under the RGEs to a somewhat larger value at the electroweak scale.

A majority of the bare-minimally constrained~\cite{Li:2011xu} parameter space of No-Scale $\cal{F}$-$SU(5)$,
as defined by consistency with the world average top-quark mass $m_{\rm t}$, the No-Scale boundary conditions,
radiative EWSB, the centrally observed WMAP7 CDM relic density limits 0.1088 $\leq \Omega h^2 \leq$ 0.1158~\cite{Komatsu:2010fb} (we assume a thermal relic), and precision LEP constraints on the lightest CP-even Higgs boson $m_{h}$ and other light SUSY chargino and neutralino mass content, remains viable.  The intersection of these experimental bounds is quite non-trivial, as the tight theoretical constraints, most notably the vanishing of $B_\mu$ at the high scale boundary, render the residual parameterization insufficient for arbitrary tuning of even isolated predictions, not to mention the union of all predictions.

\begin{figure}[htf]
        \centering
        \includegraphics[width=0.45\textwidth]{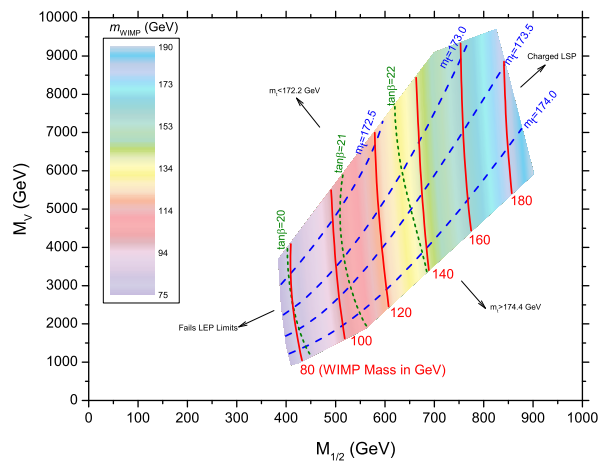}
        \caption{The bare-minimally constrained parameter space of No-Scale $\cal{F}$-$SU(5)$ is depicted as a function of the gaugino boundary mass $M_{1/2}$ and the vector-like mass $M_{\rm V}$. The WIMP mass, top quark mass $m_t$, and tan$\beta$ are demarcated via the solid, dashed, and dotted contour lines, respectively.}
        \label{fig:wimp_mass}
\end{figure}

\begin{figure}[htf]
        \centering
        \includegraphics[width=0.45\textwidth]{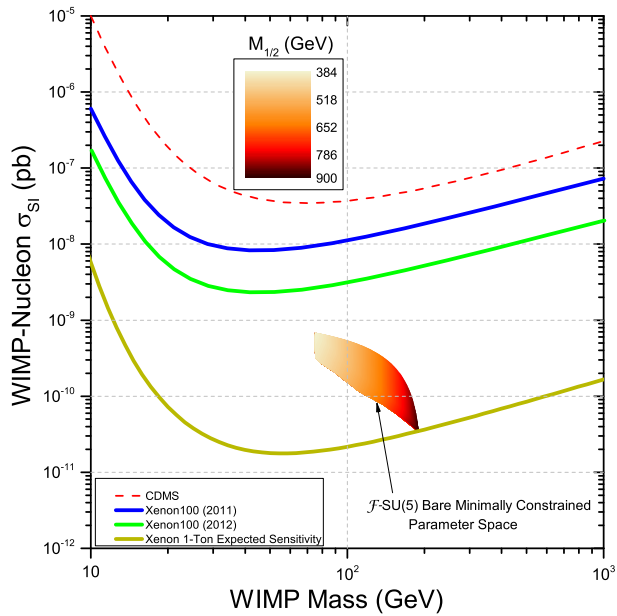}
        \caption{Direct dark matter detection diagram associating the WIMP mass with the spin-independent annihilation cross-section $\sigma_{\rm SI}$.}
        \label{fig:Xenon_sigma}
\end{figure}

The cumulative result of the application of the bare-minimal constraints shapes the parameter space into the uniquely formed profile situated in the $M_{1/2},M_{\rm V}$ plane exhibited in Fig.~(\ref{fig:wimp_mass}), from a tapered light mass region with a lower bound of $\tan \beta$ = 19.4 demanded by the LEP constraints, into a more expansive heavier region that ceases sharply with the charged stau LSP exclusion around tan$\beta \simeq$ 23, where we overlay smooth contour gradients of top quark mass, tan$\beta$, and the WIMP mass. The bare-minimal constraints set lower bounds at about $M_{1/2} \simeq$ 385 and $M_V \simeq$ 925 GeV correlated to the lower bound on tan$\beta$ of around 19.4, and upper bounds near $M_{1/2} \simeq$ 900 and $M_V \simeq$ 9.7 TeV, correlated to the upper bound on tan$\beta$ at about 23. The parameter space in Fig.~(\ref{fig:wimp_mass}) does not include the most recent constraint on the Higgs mass, though when applying the mass limits $125 \lesssim m_h \lesssim 126$, the large region of model space in Fig.~(\ref{fig:wimp_mass}) is reduced to a narrow strip of space along the lower edge of the region~\cite{Li:2012qv,Li:2012jf}.

The proportional rescaling associated with the single massive input $M_{1/2}$ explains the ability to generate the WMAP7 successfully and generically, where we assume a thermal relic. The correct DM relic density can be generated by the LSP neutralino and light stau coannihilation. All considered, it indicates how finely naturally adapted (not finely tuned) No-Scale $\cal{F}$-$SU(5)$ is with regards to the question of relic density. Although currently safe, it does appear that the full model space may be effectively probed in the near future by the extended reach of the ongoing data collection at XENON100 and the expansion to the 1-ton XENON. The relevant scale dependent sensitivity contours to spin-independent DM-nucleon scattering are depicted in Fig.~(\ref{fig:Xenon_sigma}), along with their relation to the putative $\cal{F}$-$SU(5)$ signal. In Fig.~(\ref{fig:Xenon_sigma}), we include the latest results from the XENON100 experiment released in 2012~\cite{Aprile:2012nq}, as well as the expected sensitivity for the 1-ton version of the XENON experiment.

The LHC has begun running at a beam collision energy of 8 TeV in 2012, though the analysis of these 8 TeV data observations is still progressing. However, we have completed a thorough examination of the complete 5 ${\rm fb^{-1}}$ at 7 TeV~\cite{Li:2012hm,Li:2012tr,Li:2012ix}, and discovered tantalizing correlations between the 7 TeV observations and the Monte-carlo predictions of No-Scale $\cal{F}$-$SU(5)$, particularly in the realm of the large multijet events, which as we discussed earlier, is expected to be a clear signature of $\cal{F}$-$SU(5)$ as larger amounts of data are collected.

\begin{acknowledgments}
This research was supported in part 
by the DOE grant DE-FG03-95-Er-40917 (TL and DVN),
by the Natural Science Foundation of China 
under grant numbers 10821504 and 11075194 (TL),
by the Mitchell-Heep Chair in High Energy Physics (JAM),
and by the Sam Houston State University
2011 Enhancement Research Grant program (JWW).
\end{acknowledgments}


\bibliography{bibliography}

\end{document}